\documentclass[aps,prb,twocolumn,superscriptaddress,english]{revtex4-1}

\usepackage[T1]{fontenc}
\usepackage{babel}
\usepackage{amsmath}
\usepackage{amssymb}
\usepackage{wasysym}
\usepackage{graphicx}
\usepackage{lipsum}
\usepackage{xcolor}
\usepackage{braket}
\usepackage{multirow}

\usepackage[linktocpage=true,
  colorlinks=true, 
  pdfborder={0 0 0},
  linkcolor=blue,
  citecolor=red,
  filecolor=yellow,
  urlcolor=blue,
  bookmarks,
  pdfauthor={},
]{hyperref}

\newcommand{\sh}{SH$_3$}
\newcommand{\lah}{LaH$_{10}$}
\newcommand{\yh}{YH$_{6}$}
\newcommand{\mg}{MgB$_2$}
\newcommand{\xbc}{$X$B$_3$C$_3$}
\newcommand{\SRbc}{SrRbB$_6$C$_6$}

\newcommand{\tc}{$T_\text{c}$}

\newcommand{\omlog}{$\omega_{\textmd log}$}

\newcommand{\ep}{\textit{e-ph}~}

\begin{document}

\title{High-Tc Superconductivity in doped boron-carbon clathrates}

\author{Simone di Cataldo}
\affiliation{Dipartimento di Fisica, Sapienza Universit\`a di Roma, 00185 Roma, Italy}
\affiliation{Institute of Theoretical and Computational Physics, Graz University of Technology, NAWI Graz, 8010 Graz, Austria}

\author{Shadi Qulaghasi} 
\affiliation{Dipartimento di Fisica, Sapienza Universit\`a di Roma, 00185 Roma, Italy}

\author{Giovanni B. Bachelet}
\affiliation{Dipartimento di Fisica, Sapienza Universit\`a di Roma, 00185 Roma, Italy}

\author{Lilia Boeri} \email{lilia.boeri@uniroma1.it}
\affiliation{Dipartimento di Fisica, Sapienza Universit\`a di Roma, 00185 Roma, Italy}

\date{\today}

\begin{abstract}
We report a high-throughput {\em ab-initio} study of the thermodynamic and superconducting properties of the recently synthesized \xbc\; clathrates. These compounds, in which boron and carbon form a sponge-like network of interconnected cages each enclosing a central $X$ atom, are attractive candidates to achieve high-\tc\; conventional superconductivity at ambient pressure, due to the simultaneous presence of a stiff B--C covalent network and a tunable charge reservoir, provided by the guest atom. 
Ternary compounds like CaB$_3$C$_3$, SrB$_3$C$_3$ and BaB$_3$C$_3$ are predicted to exhibit \tc $\lesssim 50 K$ at moderate or ambient pressures, which may further increase up to 77$~K$ if the original compounds are hole-doped by replacing the divalent alkaline earth
with a  monovalent alkali metal to form ordered $XY$B$_6$C$_6$ alloys.
 \end{abstract}

\pacs{~}
\maketitle

\section{Introduction}

The {\em hydride rush} initiated by the report of a superconducting critical temperature ($T_\text{c}$) of 203 K
in \sh~\cite{DrozdovEremets_Nature2015, Duan_SciRep2014} completely reshaped the
landscape of research on superconductivity, delivering the first example of room-temperature superconductivity after more than one century since its discovery,~\cite{snider2020room} and introducing a new paradigm for material prediction heavily relying on computational methods.~\cite{Oganov_review,Zurek_review,boeri2019road,pickard2020superconducting,flores2020perspective,sannacombining,Roadmap_RTS}

\begin{figure}[t]
	\includegraphics[width=1\columnwidth]{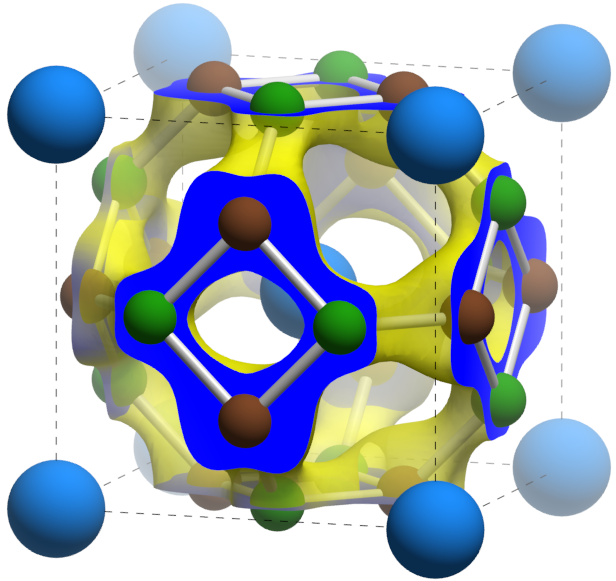}
	\caption{Crystal structure of \xbc{}. Space group:  $Pm\bar{3}n$ (223); large blue, small brown and green spheres indicate $X$, B and C atoms, in Wyckoff positions $2a$, $6d$, and $6c$, respectively. The yellow isosurface, obtained for $X$=Sr but representative of the valence-band-top wave\-function for all the \xbc\;compounds considered in this work (see also Fig.\ref{fig:bands}), closely wraps the network of B--C bonds  (see text); electric-blue lamellas mark its cross section at the cell boundaries. \label{fig:struct}
	 }	
\end{figure}

The focus of the field is gradually shifting from achieving ever larger \tc's to finding new high-\tc\; materials which can 
operate at (or close to) ambient pressure. Migdal-Eliashberg theory for superconductivity suggests that, besides hydrides,~\cite{oganov_CeH,PhysRevB.104.L020511} obvious candidates for observing high-\tc\; superconductivity 
from conventional (electron-phonon, hereafter \ep\!\!) interaction are light-element compounds. In particular, boron and/or carbon compounds combine large phonon frequencies
and covalent bonds, the two most important ingredients underlying high-\tc\; (conventional) superconductivity.~\cite{nagamatsu2001superconductivity,PRL_Pickett_2001,Boeri-diamond}
Several theoretical predictions of conceivable high-\tc\; borides and carbides date back to the early 2000's,
but none of them has been experimentally verified; \cite{rosner2002prediction,mauri_B12,kolmogorov_lib_PRB2006,moussa2008constraints,savini2010first} twenty years later,
the problem is being revisited with modern ab-initio methods for structural prediction, which permit to address the thermodynamic stability of different structures across the phase diagram.~\cite{jay2019theoretical,saha2020high}.
Last year, Zhu et al.~\cite{zhu2020_b3c3framework} reported the formation of a previously unknown boron-carbon  phase, with chemical
formula \xbc{}, which is a structural analogue of high-\tc\; sodalite clathrate hydrides, like \lah, \yh\; 
etc.~\cite{La_EXP1, La_EXP2, La_EXP3,Hclathrates_Wang_PNAS2012,Ma_PRL_2017,heil2019superconductivity} 
The structure, shown in Fig.~\ref{fig:struct}, consists of a bcc lattice of 
interconnected \textit{truncated octahedral}  boron-carbon cages, each enclosing a guest $X$ atom.

The empty B--C cage is dynamically and thermodynamically unstable at ambient pressure; 
however, a few \xbc\; compounds can be stabilized under pressure and exhibit a variety of attractive physical 
properties, such as superhardness (La) and ferroelectricity (Sc).~\cite{zhu2020_b3c3framework,zhu2020_ferroelectric,Strobel_LaB3C3,SrB3C3_SC} 
%

SrB$_3$C$_3$, which forms in diamond anvil cells at high pressure (57 GPa), but survives close to room pressure, has been theoretically predicted to superconduct below 40 K,~\cite{SrB3C3_SC} a \tc\;  close to the current record for ambient-pressure conventional superconductivity, held by
magnesium diboride (\mg). Given the extreme versatility of cage-like structures observed in superhydrides, a suitable choice of the guest atom is likely to attain even higher  \tc\;;
here we explore this possibility, by performing a detailed study of thermodynamic and superconducting properties of hypothetical B--C clathrates, using first-principles methods based on Density Functional Theory (DFT).~\cite{foot:CD}

Our sampling of conceivable compositions starts from an overall scan of the first 57 elements of the periodic table. We estimate the
dynamical and thermodynamic stability of the \xbc\; compounds with  $X=$ H-La, and find
that 
only five (Ca, Sr,Y, Ba, La) form stable \xbc\; compounds,  none of which represents a substantial improvement over \mg\; in terms of \tc\; and thermodynamic stability. 

We then consider the possibility of combining pairs of the above elements to optimize the electronic properties in the direction of a higher \tc, and find a few ordered alloys $XY$B$_6$C$_6$, where $X$ and $Y$ are mono- and di-valent elements,
which should exhibit critical temperatures above the liquid nitrogen boiling point.
 These compounds form under pressure, but may be quenched down to room pressure without losing their remarkable properties;
we argue that a further fine tuning of the doping level is very likely to give their \tc\; an extra boost.

\section{Computational Details}
\label{sect:CD}

Our calculations, based on the Density Functional Theory, were performed using \verb|Quantum ESPRESSO| (QE), \cite{QE_JPCM_2009, QE_JPCM_2017} scalar-relativistic Optimized Norm-Conserving Vanderbilt pseudopotentials,\cite{Hamann_PRB_2017_ONCV}  and a Perdew-Burke-Ernzerhof exchange-correlation functional.\cite{PhysRevLett.77.3865} The wavefunctions were expanded using a 
plane-wave basis set, with a cutoff of 80 Ry. The integration over the Brillouin zone was performed on a 6$\times$6$\times$6 $\vec{k}$-space grid, using a Methfessel-Paxton\cite{PhysRevB.40.3616} smearing width of 0.06 Ry. Such a choice ensures a 0.5 meV/atom convergence of the total energy and a 0.5 meV convergence of phonon frequencies. 

The dynamical matrices and the related \ep\!\! matrix elements were calculated using linear response theory.\cite{Baroni_RMP2001} The Brillouin zone was sampled on a 4$\times$4$\times$4 $\vec{q}$-grid (phonons), and the integration of the \ep\!\!  matrix elements was carried out on a 30$\times$30$\times$30 $\vec{k}$-grid (electrons) and a gaussian smearing width of 200 meV.  The superconducting critical temperature was estimated by numerically solving the isotropic \'{E}liashberg equations employing the \'{E}liashberg functions obtained in linear response.

\section{High-Throughput study of \xbc\; clathrate stability}

\begin{figure*}[t]
	\includegraphics[width=2.10\columnwidth]{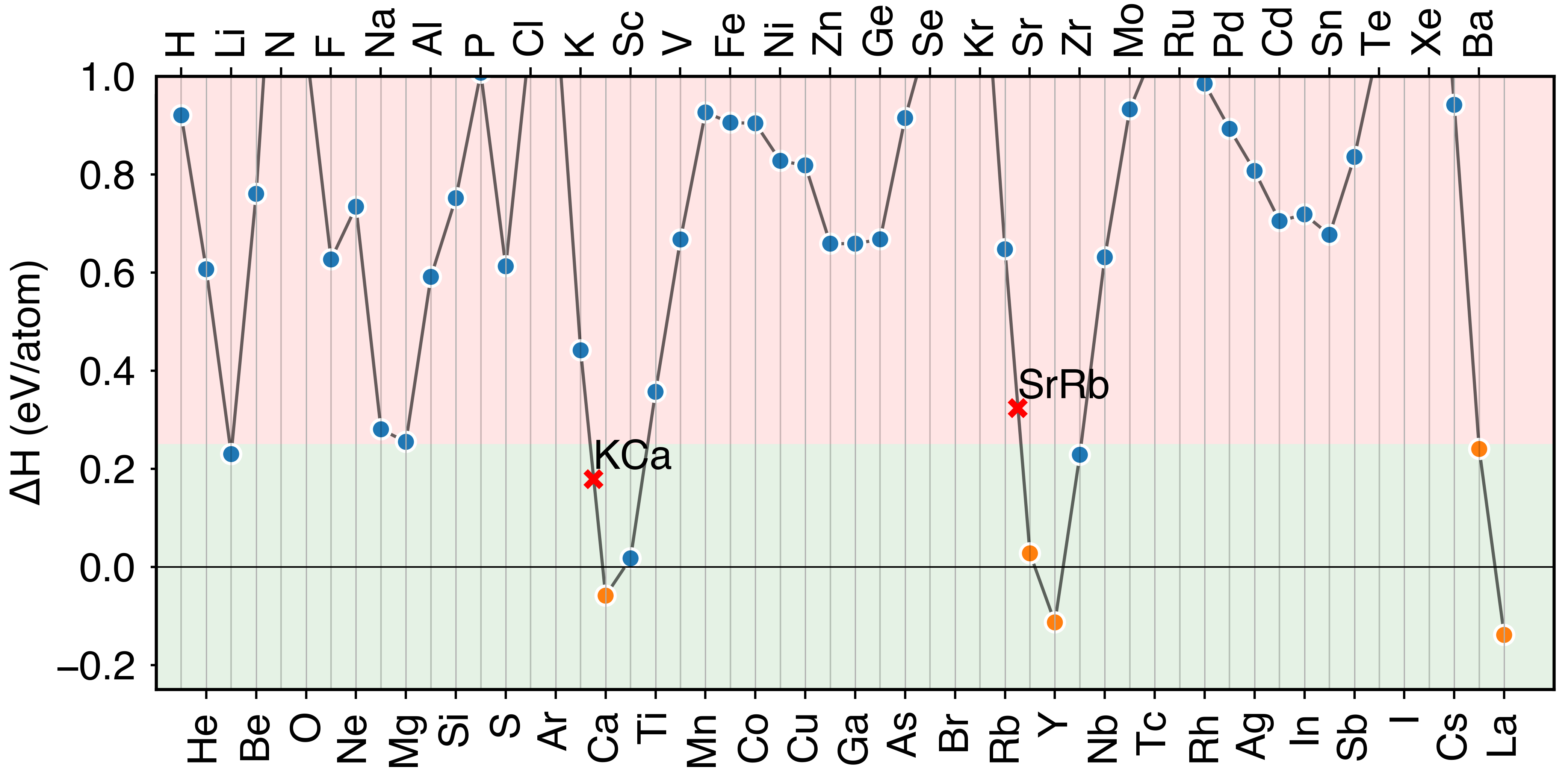}
	\caption{Decomposition enthalpy of \xbc\; compounds for the first 57 elements of the periodic table ($Z\!=\!1...57$; the chemical symbol is shown above for odd-$Z$ elements, below for even-$Z$ elements). The color of markers relates to dynamical stability: blue dots indicate dynamically stable compounds, orange dots dynamically unstable. All quantities are calculated at $P=0$.}
	\label{fig:stability}
\end{figure*}

The main results of our initial high-throughput screening are summarized in Fig.~\ref{fig:stability}.~\cite{foot:CD}
For each of the 
first 57 elements of the periodic table ($X$=H-La), we performed a full relaxation of the \xbc\; structure at ambient pressure. 
For the relaxed structures, we computed the decomposition enthalpy with respect to the ground-state elemental structures
$\Delta H = H(XB_{3}C_{3})- (H(X)+ 3H(B)+3H(C))$, where $H(XB_{3}C_{3})$ is the enthalpy of the \xbc\; structure relaxed at a given pressure, and
$H(X)$, $H(B)$ and $H(C)$ are the enthalpies of the three elements in their ground-state structure
at the same pressure.\cite{foot:GS}
This quantity is shown on a per/atom basis in Fig.~\ref{fig:stability}; a comparison between $\Delta H(P = 0\; GPa)$ and $\Delta H(P =\; 50 GPa)$ is shown in Fig. S2 of the Supplemental Material \cite{SM_link}.  
 
Note that the decomposition enthalpy $\Delta H$ thus defined provides only a lower bound on the actual formation enthalpy. The latter also includes the decomposition into all possible binary and ternary phases on the hull, which is unfeasible 
 in a high-throughput study.
  However, a negative $\Delta H (P)$ usually indicates that the phase will survive down to pressure $P$, if formed at higher pressure.
  
 In our scan, only Ca, Y, and La are predicted to remain stable down to ambient pressure; and,
indeed, LaB$_3$C$_3$ has been experimentally reported to form at 1 atm.\cite{Strobel_LaB3C3}

The green-shaded region in Fig.\ref{fig:stability} indicates the range of {\em metastability}, i.e. an energy region with $\Delta H > 0$, which
encloses structures that may be synthesized under appropriate experimental conditions.
Although there is no rigorous criterion for metastability, values of $\Delta H$ 
ranging from a few tens to a few hundreds meV are
considered acceptable in literature;~\cite{aykol2018thermodynamic}
here we choose $\Delta H = 250$ meV,  roughly corresponding 
to the average $pV$ term at 50 GPa.

In addition to thermodynamic stability, Fig.\ref{fig:stability} includes information on the {\em dynamical} (phonon) stability of 
\xbc{}; compounds, estimated from linear-response calculations on a $2^{3}$ grid in reciprocal space.~\cite{foot:CD} Compounds indicated with orange symbols have real (positive) frequencies on all points of the grid, while those with blue (empty) symbols have at least one negative (imaginary) optical frequency, and hence are dynamically unstable. 
Our criterion correctly classifies ferroelectric 
ScB$_3$C$_3$ as dynamically unstable.~\cite{zhu2020_ferroelectric}

A first inspection of the figure reveals that the formation of \xbc\; structures is energetically unfavorable for most elements in the periodic table; 
only three elements have negative formation enthalpies. Interestingly, all three compounds are
predicted to be also {\em dynamically} stable, which is a strong indication that they may be synthesised, under appropriate
conditions. In addition to the thermodynamically stable compounds at zero pressure,
two more compounds,
 SrB$_3$C$_3$ and BaB$_3$C$_3$, which have a positive decomposition
 enthalpy at  ambient pressure, can be stabilized at higher pressure.
The decomposition pressures ($P_d$), estimated by linear interpolation of the decomposition
enthalpies between zero and 50 GPa, are 5 and 30 GPa in Sr and Ba, respectively.
The corresponding formation pressures are expected to be significantly higher: for SrB$_3$C$_3$,
where experimental data exist, the difference is $\sim 50$ GPa. Assuming a similar
shift, the formation pressure of BaB$_3$C$_3$ could be as high as 80 GPa.

The five elements which form stable \xbc\; compounds (i.e. Ca, Sr, Y, Ba, La) all belong   either to the II A or to the III A group of the periodic table. An inspection of their electronic structure and of their structural data reveals that these elements lie in a sweet spot 
of valence and atomic radius; moreover, their low electronegativity implies that, once within the B--C cage, they are completely ionized and their
valence states do not mix with B--C states, and thus do not disrupt the stability of the structure. The dependence of the parameter on the $X$ atom is shown in Fig. S1 of the Supplemental Material \cite{SM_link}. 

\section{Electronic Structure}

\begin{figure}[t]
	\includegraphics[width=1.00\columnwidth]{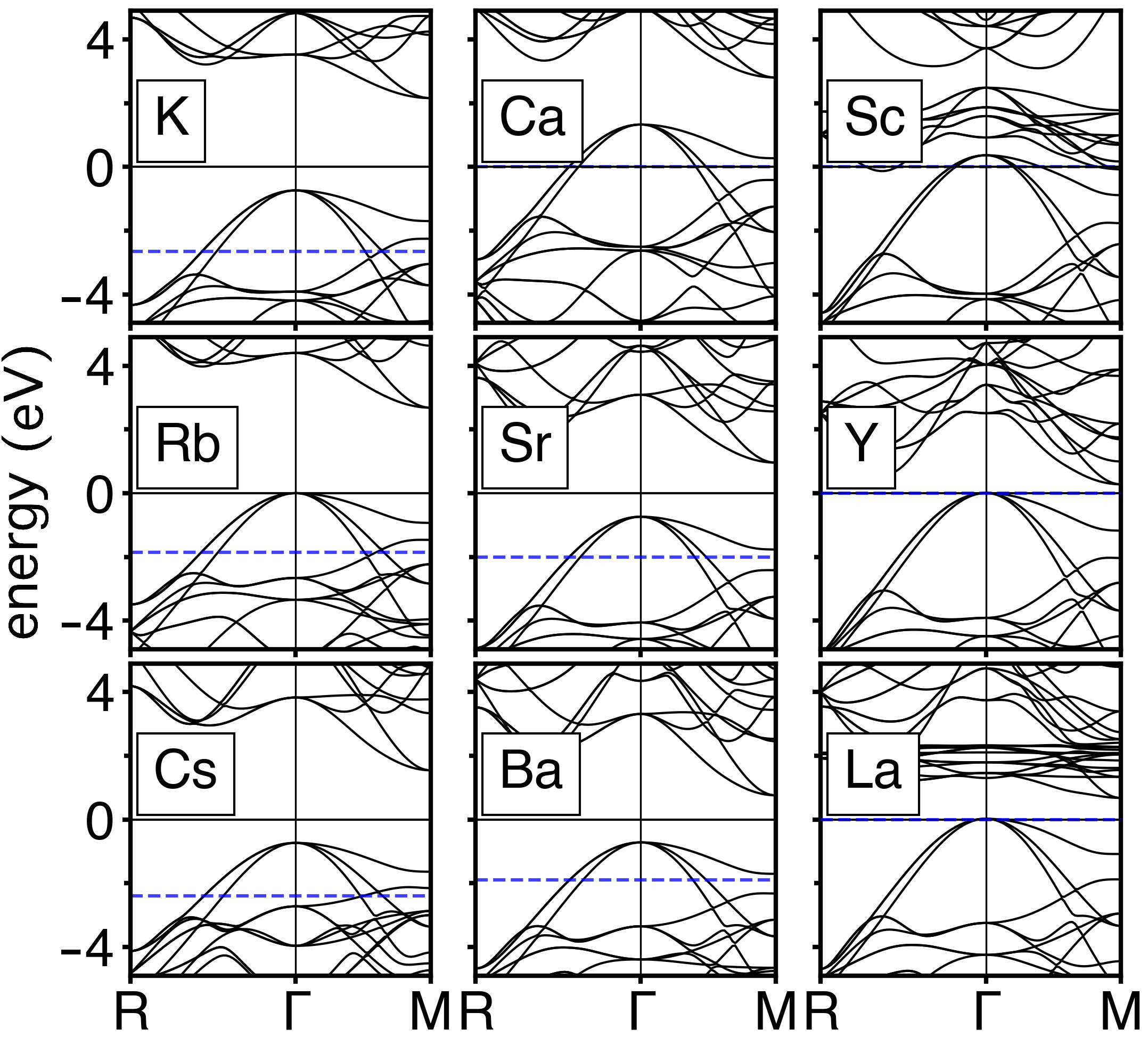}
	\caption{Band structure of the nine  \xbc\; compounds considered in this work. The energy zero is set equal to the valence band tops, the Fermi energy is a blue dashed line.}
	\label{fig:bands}
\end{figure}

Fig.~\ref{fig:bands} shows an enlargement of the electronic structure in the Fermi level region for the five dynamically stable compounds at ambient pressure, and their immediate neighbors in the periodic table (K, Rb, Cs), arranged by period. 
For symmetry, we also include scandium, although dynamically unstable (ferroelectric\cite{zhu2020_ferroelectric}).
For each compound the energy zero is chosen in such a way as to coincide with the top of its valence bands;  the position of its
Fermi level is shown as a blue dashed line.  
Such a choice emphasizes the fact that the electronic bands of \xbc\; compounds with $X$ belonging to the same period essentially follow a rigid-band model:
in the valence region they are of almost pure B--C character; the guest $X$ atoms, almost completely ionized, donate their charge to the B--C sublattice, filling its bands.

\xbc{} with the tri-valent guest elements Y and La are insulating: the 24 valence electrons are, in fact, sufficient to fill up the 12 B--C bonding bands,
separated by a $\sim $~1~eV gap from the conduction bands, where antibonding B--C and $X$ guest states do mix. 
Not so with Sc, whose $d$ states, lower in energy, partially hybridize with the top of the valence band too, resulting in a poor metal (upper right panel of Fig.\ref{fig:bands}).

Replacing tri- with di- and mono-valent elements (middle and left panels of Fig.\ref{fig:bands}, respectively), the Fermi level is shifted into the valence band, effectively doping it with holes.
In Fig.~\ref{fig:struct} we show the $X$B$_3$C$_3$ crystal structure with an isocontour at one fourth of the maximum value of the squared modulus of the wavefunction at the top of the valence bands \cite{foot_degen_bands}
The yellow isosurface, obtained for $X$=Sr but representative of the valence-band-top wave\-function for all the \xbc{} compounds appearing in Fig.\ref{fig:bands}, closely wraps the network of B--C bonds.
Holes doped into this band will experience an extremely strong \ep\; coupling to bond-stretching phonons; a similar
mechanism is at the heart of the remarkable superconducting properties
 of other B--C compounds, such as \mg, boron-doped diamond, graphane.~\cite{nagamatsu2001superconductivity,ekimov2004superconductivity,Boeri-diamond,moussa2008constraints,savini2010first,saha2020high}
In \xbc, this very mechanism leads to substantial \tc's. For metallic structures, 
using linear-response \ep\!\!  spectra and  the McMillan-Allen-Dynes formula, assuming $\mu^*=0.1$,
we predict \tc\; values between 40 and 50 K -- see table I.
For Sr and Ca, calculations are performed at zero pressure, where the two compounds are dynamically stable; 
Ca has a very soft mode along the $\Gamma-M$ line. For Ba, which is dynamically unstable
at ambient pressure, we report calculations at 30 GPa, where the compound is dynamically stable.
In all cases, \ep\!\!  spectra are qualitatively very similar to
those shown  for the doped samples in Fig.~\ref{fig:phonons} -- See Figs. S3 to S8 of the Supplemental Material \cite{SM_link};
phonon frequencies extend up to $\sim 100$~meV; the \ep\!\! 
coupling is spread  over a wide range of frequencies, with a strong enhancement in the mid-frequency region where softer bond-stretching
phonons are concentrated.

\begin{table}[htpb]
\begin{tabular}{c|c|c|c|c|c}	
    \hline
    \hline
    
                               &      P$_d$ &   $\Delta$H$_{P=0}$   &    \omlog            &  $\lambda$       & \tc              \\
                               &           (GPa)            &         (meV/atom)        &      (meV)          &                        &        ( K )     \\
 \hline
      Ca                     &         0                &          -91                 &         17.5$^{*}$    &     1.9          &       48                \\
       Sr                     &         0                 &        +5                    &        51.0             &     1.0          &       44              \\
       Ba                    &        30                &        +203               &         49.8            &     1.1          &        50              \\
    \hline
    \hline
\end{tabular}
\caption{\label{table1}
Calculated superconducting properties of stable \xbc\; compounds; the asterisk indicates that CaB$_3$C$_3$ experiences a remarked softening of one of the optical modes between $M$ and $\Gamma$.
}
\end{table}

\section{Doping}

\begin{figure}[t]
	\includegraphics[width=0.95\columnwidth]{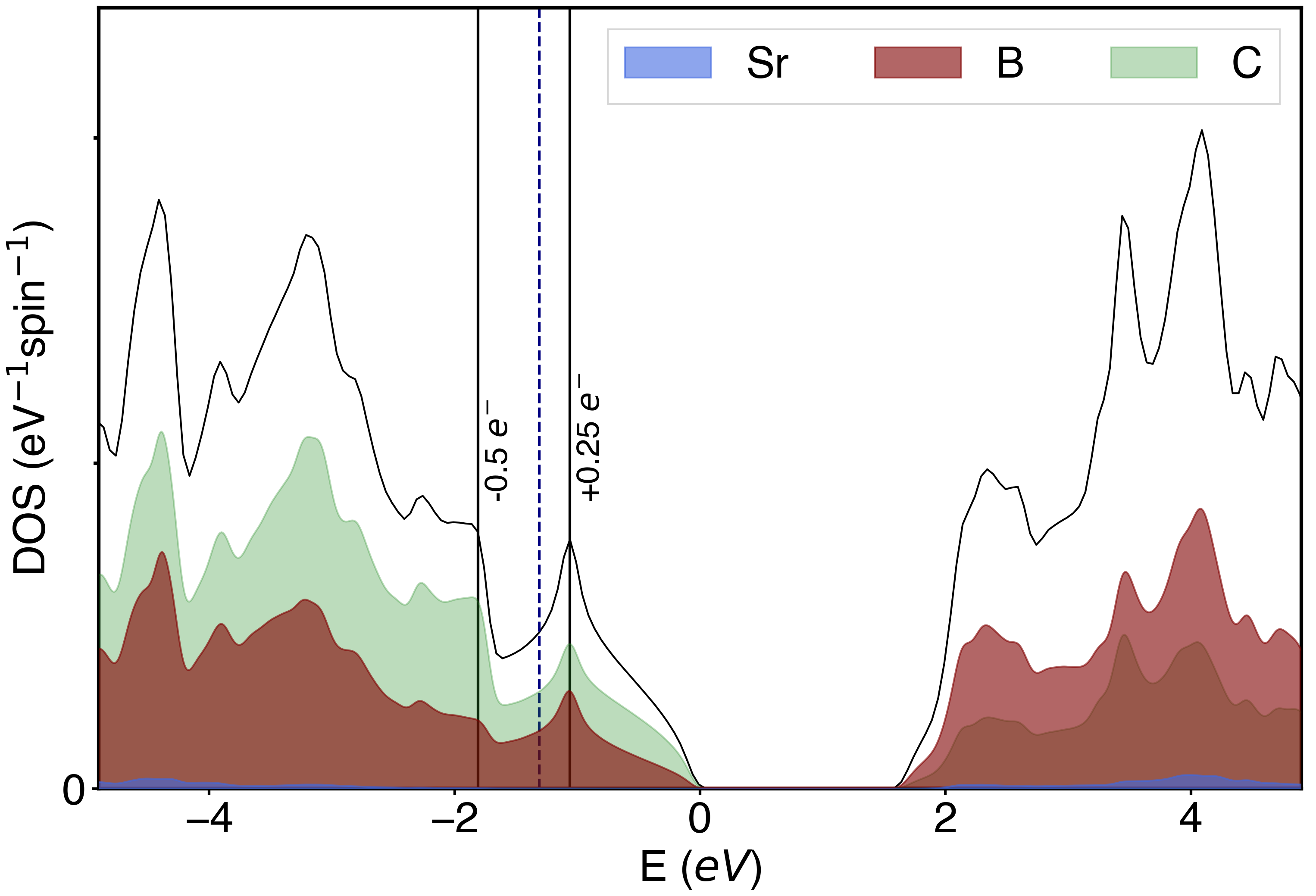}
	\caption{Electronic Density of States of SrB$_3$C$_3$ in the vicinity of the Fermi level
	(vertical dashed line). See text.}
	\label{fig:DOSRB}
\end{figure}

The rigid-band behavior  observed in Fig.~\ref{fig:bands}
suggests that a partial substitution at the guest site may be used to improve the
superconducting properties of alkaline-earth BC-clathrates.
Fig.~\ref{fig:DOSRB} shows an enlargement of the electronic Density of States (DOS) of SrB$_3$C$_3$ in the region around the Fermi level.
It lies below the valence band top, within a $\sim 1$eV--wide pseudogap
which clearly separates a DOS shoulder, corresponding to a 1/2 hole/f.u. doping, from an equally tall sharp peak, corresponding to a 1/4 electron doping.
While 1/4 electron doping requires a large supercell,  1/2 hole doping may be easily simulated by a simple cubic supercell, in which half of the divalent Sr atoms are replaced by monovalent Rb.
At ambient pressure, such an ordered alloy is 12 meV/atom more stable than isolated SrB$_3$C$_3$ + isolated RbB$_3$C$_3$ , 
but its decomposition enthalpy with respect to the pure elements is high: $\Delta H$=305 meV/atom. In analogy to SrB$_3$C$_3$, 
we expect that, at higher pressure, $\Delta H$ be reduced and \SRbc become stable. Following this line of reasoning, the estimated decomposition pressure would be $P_d=$ 42 GPa.

For KCaB$_6$C$_6$ and CsBaB$_6$C$_6$, obtained by doping CaB$_3$C$_3$ and BaB$_3$C$_3$ with K and Cs, respectively,
the predicted decomposition pressures are 18 and 117 GPa -- see table \ref{table2}.
Hence, a correlation seems to emerge between the average atomic radius and the decomposition pressure: smaller atoms require lower pressure.


\begin{figure}[t]
	\includegraphics[width=1.00\columnwidth]{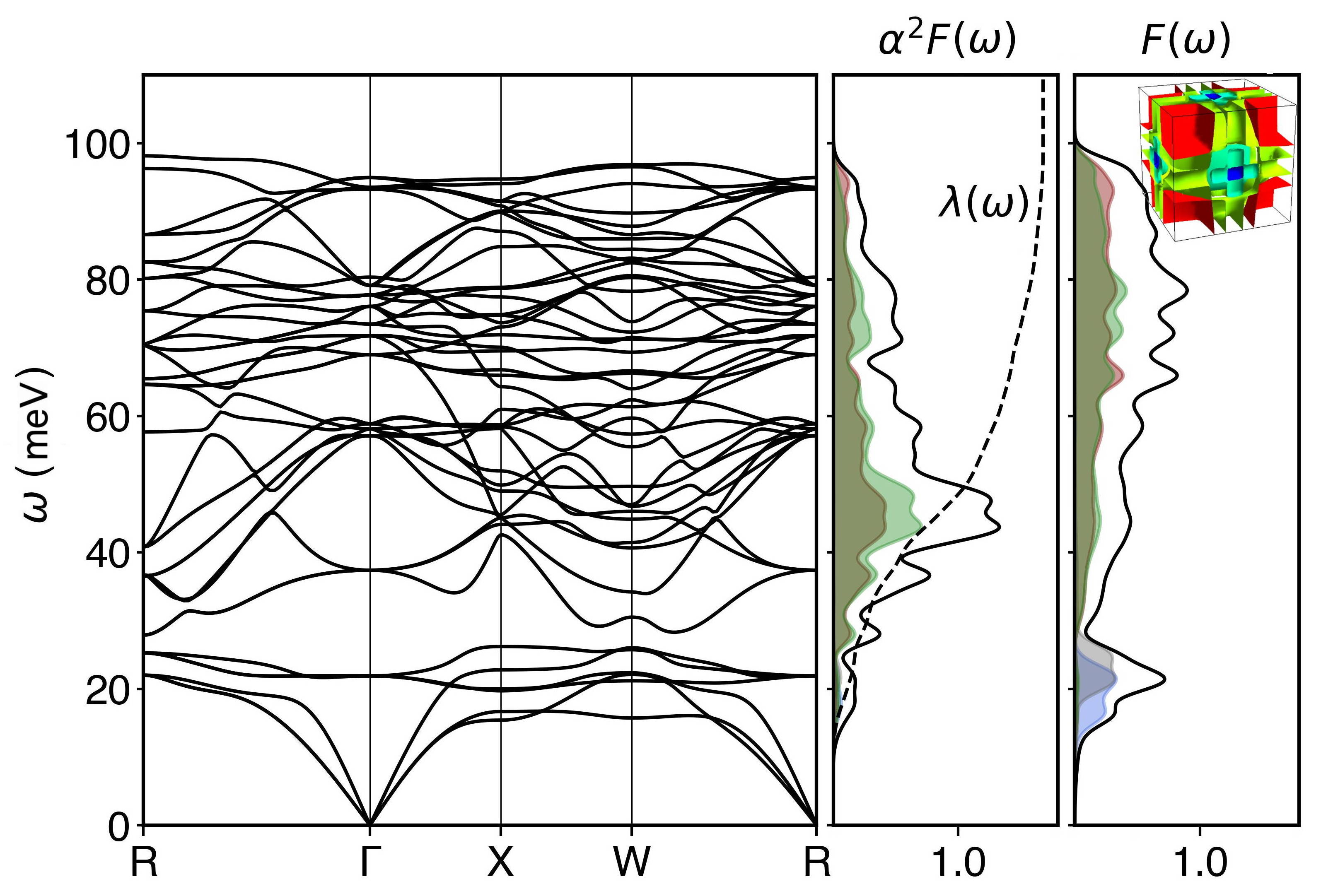}
	\caption{Phonon Dispersions (left), Density of States (middle), Eliashberg spectral function and Fermi surface (right) of SrRbB$_6$C$_6$ (\tc = 77 K). }
	\label{fig:phonons}
\end{figure}

 Fig.~\ref{fig:phonons} shows the calculated phonon 
dispersions and Density of States of \SRbc, together with the \ep\!\!  Eliashberg spectral function $\alpha^2 F(\omega)$.
The corresponding figures for KCaB$_6$C$_6$ and CsBaB$_6$C$_6$ are shown in the Supplemental Material \cite{SM_link}.
The \ep\!\! coupling is mostly distributed on modes of the B--C host sublattice; a strong enhancement 
in the mid-frequency region, where soft bond-stretching phonons are concentrated, translates into 
a relatively low \omlog\;($\sim 44$~meV) and a very large total \ep\;coupling constant $\lambda$=1.7.
Given the strong-coupling regime, we estimated the superconducting critical temperature by
solving the full (isotropic) Migdal-Eliashberg equations. With a constant  $\mu^*=0.1$ 
we obtain a \tc\;\!\! of 77 K.

Table \ref{table2} reports the relevant superconducting parameters for the other two ordered alloys considered here. The quantities calculated at ambient pressure for KCaB$_6$C$_6$ are extremely close to those estimated for
\SRbc. For CsBaB$_6$C$_6$, which turns out to be weakly dynamically unstable at ambient pressure, we report results at 10 GPa;
at this pressure, the predicted \tc\;\!=82 K is even slightly larger than the liquid nitrogen boiling point.
These results indicate that, while the atomic radius has a major impact on pressure, its effect    on \tc~is almost negligible.

\begin{table}[htpb]
\begin{tabular}{c|c|c|c|c|c}	
    \hline
    \hline
                               &     $P$ ($\Delta$H=0)   &   $\Delta$H ($P$=0)   &    \omlog &  $\lambda$ & \tc    \\
                               &         (GPa)               &         (meV/atom)        &      (meV)             &                   &        ( K )            \\
    \hline

KCa         &    18                      &         +160                &         43.9           &     1.6         &       77              \\
SrRb       &           42              &         +305                &          44.0           &    1.7          &       78             \\
CsBa     &           117              &        +556                &          35.4*          &     2.2*          &       82*             \\
\hline
B-diam \cite{saha2020high} & -- & +440 & 41.0 & 2.3 & 75 \\
    \hline
    \hline
\end{tabular}
\caption{\label{table2}
Predicted thermodynamic and superconducting properties of ordered alloys $XY$B$_6$C$_6$;  quantities are defined as in table \ref{table1}. The asterisk reminds that for CsBa the superconducting parameters were calculated at $P\!\!=\!10$ GPa (see text). In the last row we report data for 50 $\%$ B-doped diamond, from Ref.~\onlinecite{saha2020high}. Note that, in that paper, \tc~was estimated by the Mc-Millan-Allen-Dynes formula, while in the present work we employ the full solution of Migdal-Eliashberg equations, which is more accurate at strong coupling.}
\end{table}


The data reported in table \ref{table2} suggest that doped \xbc\; compounds represent a substantial step forward 
in our understanding of conventional superconductors.
Critical temperatures close to or larger than nitrogen boiling point (77 K) at room pressure have never
been reported in any conventional superconductor so far; indeed, the current \tc\; record is much lower -- 39 K in \mg.
On the last row of the table, we reported the corresponding data for heavily (50 $\%$)
B-doped diamond, which has also been predicted to exhibit \tc's close to 80 K.~\cite{moussa2008constraints,saha2020high} 
It is remarkable to observe that the main superconducting parameters are extremely close, indicating
similar lattice stiffness, bonding properties, nature and distribution of the \ep\; coupling between
diamond and \xbc{} structures.
The key difference is that $\Delta H$, in heavily B-doped diamond, is so high ($\sim 440$ meV)
that its experimental synthesis is likely impossible -- the highest reported B- doping levels in diamond are 4-5 times smaller --
while, according to  our calculations, KCaB$_3$C$_3$, similarly to SrB$_3$C$_3$, could more easily form at moderate pressures, and then, once formed, it may be quenched down to ambient pressure, without spoiling its substantial \tc. 
In addition, the empirical correlation between average atomic radii and stabilization pressure, the quasi-rigid-band behavior (Fig.\ref{fig:bands}) and the steep energy dependence of the DOS at both edges of the pseudogap (Fig.\ref{fig:DOSRB}), suggest that both the synthesis conditions and the superconducting properties may be further improved by adjusting the doping level and/or combining atoms with different atomic radii.

\section{Conclusions}

In conclusion, in this work we performed a broad-range study of the thermodynamic stability and superconducting properties of $X$-doped B--C clathrates \xbc{}. These compounds, recently synthesized under high pressure, are structurally related to record cage-like superhydrides like LaH$_{10}$ and YH$_{6}$, while their bonding and electronic properties, determined by the covalent B--C sublattice, are analogous to those of borides and carbides like \mg{}; and doped diamond. ~\cite{nagamatsu2001superconductivity,ekimov2004superconductivity}

Based on a full scan of the periodic table, we find that only five elements (Ca, Sr, Y, Ba and La) form stable \xbc{} compounds at ambient or moderate pressure. Two of them (La, Sr) have already been experimentally synthesized.\cite{zhu2020_b3c3framework,Strobel_LaB3C3,SrB3C3_SC} While the superconducting \tc{} of ternary \xbc{} compounds are comparable to those of \mg{}, the best conventional superconductor known so far,  ordered alloys $XY$B$_6$C$_6$ containing mono- and divalent elements are, instead, expected to yield a substantial improvement, reaching \tc{}'s as high as 77 K. 
KCaB$_6$C$_6$, where a high \tc{}; cohexists with a relatively moderate stabilization pressure (18 GPa), represents the most promising combination of elements among those  considered here. A careful tuning of dopants and compositions may further increase \tc\; and/or reduce stabilization pressure.

\textbf{Note:} While writing this manuscript, we became aware of another study on superconductivity in doped \xbc{}; compounds; its main results are in excellent agreement with ours.~\cite{AB3C3_doping}

\begin{acknowledgments}
The authors acknowledge support from Fondo Ateneo-Sapienza 2017-2019. The computational resources were provided by CINECA through project IsC90-HTS-TECH\_C, the Vienna Scientific Cluster (VSC) through Project P30269-N36 (Superhydra), and the dCluster of the Graz University of Technology. We thank Antonio Sanna for kindly sharing the code to numerically solve the isotropic \'{E}liashberg equations

\end{acknowledgments}
\clearpage


\begin{thebibliography}{45}
\bibitem [{foo({\natexlab{b}})}]{foot:GS}%
	\BibitemOpen
	\href@noop {} {}  \bibinfo {note} {For all elements, we
		considered the ground-state structures under pressure, reported in
		literature; when the structures were not available, we employed structures
		predicted by evolutionary algorithsm (USPEX).}\BibitemShut {Stop}%
\bibitem [{foo({\natexlab{c}})}]{foot_degen_bands}%
	\BibitemOpen
	\href@noop {} {}\ \bibinfo {note} {Specifically, since the
		dispersion relationship at the top of the valence band is threefold
		degenerate, we used the average of the squared modulus of the wavefunctions
		for the degenerate manifold.}\BibitemShut {Stop}%
\end{thebibliography}

%

\end{document}